\documentclass[aps,prl,twocolumn, superscriptaddress, showkeys]{revtex4}

\usepackage{graphicx}
\usepackage{url}
\usepackage{color}
\usepackage{amsfonts}
\usepackage{amsmath}
\usepackage{dsfont}
\usepackage{float}
\usepackage{setspace}
\usepackage{wrapfig}

\newcommand{\dd}{\mbox{d}}

\renewcommand{\S}{S}
\newcommand{\s}{s}

\renewcommand{\(}{\left(}
\renewcommand{\)}{\right)}
\newcommand{\eqsize}{\footnotesize}

\newcommand{\myscalerFig}{.97}

\begin{document}


\title{An analytically tractable model of neural population activity in the presence of common input explains higher-order correlations and entropy}


\author{Jakob H Macke}
\email{jakob@gatsby.ucl.ac.uk}
\affiliation{Gatsby Computational Neuroscience Unit, University College London and University of T\"ubingen, Germany}
\author{Manfred Opper}
\affiliation{Artificial Intelligence Group, Technical University Berlin, Germany}

\author{Matthias Bethge}
\affiliation{Werner Reichhardt Centre for Integrative Neuroscience, Bernstein Centre for Computational Neuroscience, MPI for Biological Cybernetics and University of T\"ubingen, Germany}

\date{\today}
\begin{abstract}

Simultaneously recorded neurons exhibit correlations whose underlying causes are not known. Here, we use a population of threshold neurons receiving correlated inputs to model neural population recordings.  We show analytically that small changes in second-order correlations can lead to large changes in higher correlations, and that these higher-order correlations have a strong impact on the entropy, sparsity and statistical heat capacity of the population. Remarkably, our findings for this simple model may explain a couple of surprising effects recently observed in neural population recordings.
\end{abstract}

\maketitle

Finding 
models for capturing the statistical structure of firing patterns distributed across multiple neurons is a major challenge in sensory neuroscience.  Recently, the Ising model \cite{Parisi_98}, originally introduced to understand ferromagnetism, has become popular for studying neural population recordings \cite{Schneidman_Berry_06,Shlens-et-al_06, Ohiorhenuan_Mechler_10}. The use of the Ising model for neural data analysis originates from the fact that it constitutes the optimum with respect to the maximum entropy (MaxEnt) rationale \cite{Jaynes_57}, and thus that deviations from the model are diagnostic of higher-order interactions, often referred to as higher-order correlations ('hocs')\cite{Watanabe_60}.   It has been argued that hocs in spike trains play a critical role for the underlying population code. They have been shown to be stimulus- and scale dependent, and to affect the sparsity of the population response \cite{Ohiorhenuan_Mechler_10}. Studies using MaxEnt models have also raised the question of how the joint entropy \cite{Schneidman_Berry_06, Roudi_Nirenberg_09} and the statistical heat capacity \cite{Tkacik_Schneidman_09} of neural populations  or natural stimuli \cite{Stephens_Mora_08} scale with population size.

Here, we provide a parsimonious, tractable population model which can account for this multitude of empirical observations. We study the effect of hocs in a phenomenological  population model with neurons receiving common input.   In our model, correlations between binary neurons are thought to arise from common Gaussian inputs into threshold neurons, and it is thus equivalent to the \emph{Dichotomized Gaussian} distribution (DG) \cite{Cox_Wermuth_02, Macke_Berens_09}. We show that the  statistical properties of the model could provide an explanation for some recent experimental observations in population recordings. Importantly, we find that magnitude of hocs in the DG is strongly modulated by pairwise correlations, and in a manner which is consistent with neural recordings. In addition, we investigate the asymptotic scaling of the entropy in the DG and MaxEnt models, and  show the impact of hocs on the sparsity of the population.  Finally, we find that  the specific heat of a population is strongly affected by hocs: It diverges with population size for models with  all-to-all correlations beyond second order, and therefore any such model will have have a critical point at unit-temperature. 

\begin{figure}[t]
\includegraphics[width=\myscalerFig\linewidth]{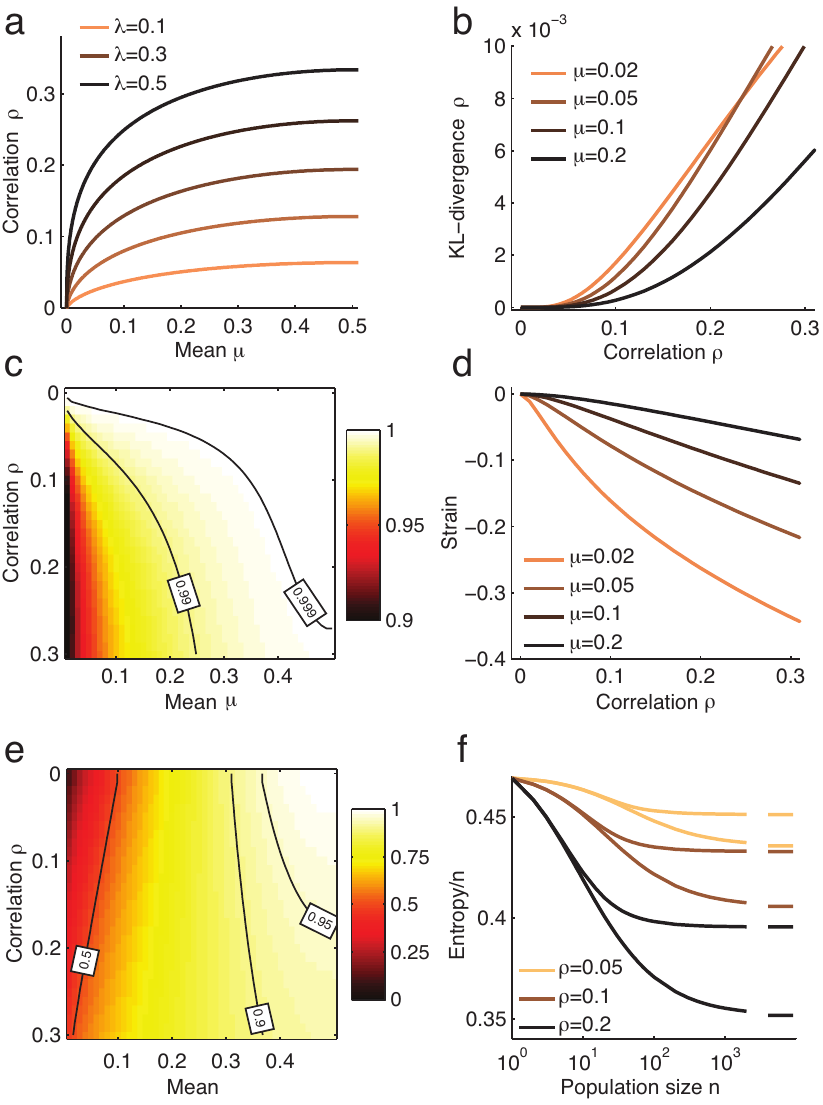}
\caption{Correlations in the DG {\bf a)} Correlations increase with firing probability $\mu$ for constant input correlation $\lambda$.  {\bf b)} KL divergence  $\Delta_{hoc}$ depends on mean firing rate $\mu$ and correlation $\rho$ in a population of size $n=5$.  {\bf c)} For $n=5$, the multi-information explained ($I_2$) by a DG is very large. {\bf d)} The strain of the homogeneous DG is negative and correlation-dependent. {\bf e)} Asymptotically, $I_2$ between the models can be very low for small correlations. {\bf f)} Scaling of the entropies of MaxEnt/DG as a function of population size $n$ for mean $\mu=0.1$, and comparison with asymptotic rates. The entropy per neuron drops initially before settling to the asymptotic value. For weak correlations, differences between models only become substantial for large $n$.\label{fig:entropycomp}} 
\end{figure}

\paragraph{The Dichotomized Gaussian is a model of correlated input.}
We model a population of $n$ binary neurons $X_i$, where a neuron is said to spike $(X_i=1)$ if its input is positive, and to be silent $(X_i=0)$ otherwise. The inputs are modelled by a correlated Gaussian with mean $\gamma$ and covariance $\Lambda$. For the outputs $X$ to have mean $\mu$ and covariance $\Sigma$, we choose $\gamma$ and $\Lambda$ such that $\Lambda_{ii}=1$,  $\mu_i=\Phi(\gamma_i)$ and  $\Sigma_{ij}=\Phi_2(\gamma_i, \gamma_j,\Lambda_{ij})-\Phi(\gamma_i)\Phi(\gamma_j)$, where $\Phi(.)$ is the cumulative distribution function (cdf) of a univariate Gaussian, and $\Phi_2(., ., \lambda)$ the cdf of a bivariate Gaussian with correlation coefficient $\lambda$. The equations above have a unique solution for any admissible moments, and can be solved numerically \cite{Macke_Berens_09}. In the special case of $\mu_i=\mu_j=1/2$,   $\Lambda_{ij}=\sin(2\pi \Sigma_{ij})$. Fig. \ref{fig:entropycomp} a shows that, for fixed input correlation and firing probability, there is a characteristic relationship between correlations and firing probabilities which is similar to that found in neural recordings \cite{Greenberg_Houweling_08}.   For analytical tractability, we here focus on homogeneous populations, i.~e. $\mu_i=\mu$ and $\Sigma_{ij}=\sigma, \Lambda_{ij}=\lambda~\forall (i \neq j)$ \cite{Parisi_98, Bohte_Spekreijse_00, Amari_Nakahara_03}. We define the pairwise correlation coefficient $\rho=\sigma/(\mu(1-\mu))$. By symmetry, all patterns $x$ with the same number of spikes are equally likely, and thus the model is fully specified by the distribution over spike counts $K=\sum_i X_i$.  

\paragraph{The effect of hocs is modulated by pairwise correlations.}
We want to determine how much additional redundancy between neurons is induced by the hocs of the correlated input model. We define $\S_{DG}$ to be the entropy of the full model, $\S_{q}$ of the MaxEnt model with interactions of order $q$, as well as $\Delta_{2}=\S_1-S_2$ and $\Delta_{hoc}=S_2-\S_{DG}$ to be the reduction in entropy due to second-- and higher-order correlations. Importantly, $\Delta_{hoc}$   corresponds to the Kullback-Leilber (KL) divergence, i.e. the expected 
log-likelihood ratio per sample between a model and its second-order approximation \cite{Grunwald_Dawid_04},  a popular measure of the magnitude of hocs  in neural recordings \cite{Schneidman_Berry_06, Ohiorhenuan_Mechler_10}. 

Figure \ref{fig:entropycomp} b shows $\Delta_{hoc}$ for a population model of size $n=5$.  Notably, small changes in firing probabilities and pairwise correlations can result in large changes in  $\Delta_{hoc}$. For example, a change of correlation coefficient from $0.05$ to $0.1$ for $\mu=0.1$ leads to an increase of $\Delta_{hoc}$ by a factor of $10.3$ (from $6.6$ to $68 \cdot 10^{-5}$). This constitutes a possible quantitative explanation for the  interesting phenomenon that hocs are much more pronounced amongst nearby cortical neurons \cite{Ohiorhenuan_Mechler_10}, for which also pairwise correlations are expected to be higher. It is also consistent with the finding that $\Delta_{hoc}$ is small  in retinal recordings with weak correlations \cite{Schneidman_Berry_06, Roudi_Nirenberg_09}. Similarly, the 'multi-information explained' \cite{Schneidman_Berry_06} $I_2=\Delta_{2}/(\Delta_{2}+\Delta_{hoc})$ of a $DG$ is large, e.g. $I_2=0.987$ for $\mu=\rho=0.1$ \cite{Roudi_Nirenberg_09}.

 We also find that the \emph{strain} \cite{Ohiorhenuan_Victor_10} of the DG-model, a measure of how much more likely a spike-triplet is as a consequence of third-order correlations, is negative ($-0.04$ for $\mu=\rho=0.1$,  using $\log_2$), and decreases with increasing correlation coefficients (Fig. \ref{fig:entropycomp} d). This is consistent with experimental observations \cite{Ohiorhenuan_Victor_10} and surprising, as it has been sugested that  a common-input model would have a higher occurrence of spike-triplets, and thus have positive  strain which increases with correlations \cite{Ohiorhenuan_Victor_10}.  Further simulations with heterogeneous correlations in the DG show that its strain is usually negative when all three pairwise correlations have the same sign. Thus, these statistical properties of our common input model are consistent with those observed in small neural populations.  

\paragraph{For large populations, $\Delta_2$ and $\Delta_{hoc}$ scale linearly with population size.}
We are interested in the scaling of the entropies of the two models with population size.
For the DG, the asymptotic probability density of the normalized counts $R=K/n$, which we denote by $f(r)$, $r \in (0,1)$ is given by
\footnote{This distribution can be derived using the method of steepest descent \cite{Amari_Nakahara_03} or by finding the likelihood of an input which has probability $r$ of inducing a spike.}: 
 \begin{align} \eqsize
f_{DG}(r)
&= \frac{1}{Z_{DG}}\exp
-\frac{1}{2}\frac{\(\Phi^{-1}(r)-\frac{\gamma\sqrt{1-\lambda}}{(1-2\lambda)}\)^2}{\lambda/(1-2\lambda)}
 \label{eq:DGasympdistrib} 
\end{align}

We can calculate the asymptotic entropy rate of the DG, $\s_{DG}=\lim_{n\rightarrow \infty}S_{DG}/n$ by decomposing it into the entropy of the spike count and the entropy conditional on the spike count, $\S(X)=\S(X|K)+\S(K)$. We note that $\S(K)$ is bounded above by $\log_2 n$, and that $\S(X|K=k)=\log_2{n\choose k}$. Using the identity $\log_2 {n \choose nr}/n\rightarrow -(r\log_2(r)+(1-r)\log_2(1-r))=:\eta_2(r)$, we can see that entropy in this model with all-to-all correlations is extensive, i.e. does not saturate, but rather scales linearly with population size for large $n$ \cite{Schneidman_Berry_06,Roudi_Nirenberg_09} with rate $\s_{DG}=\int_{0}^{1} f_{DG}(r) \eta_2(r) \dd r \label{eq:DGasympentropy}$.

We calculate the maximal entropy for large $n$ by finding the spike count distribution $P_{isi}(k)$ which maximizes $H(X|K)$. 
The solution of this constrained linear optimization problem is a  mixture of two delta peaks,  $f_{isi}(r)=p_{1}\delta(r-r_{1})+p_{2} \delta(r-r_{2}) $ with locations $r_{1,2}=1/2\pm\sqrt{1/4-\mu+\mu^2+\sigma}$ 
\footnote{This solution can be verified using the Karush-Kuhn-Tucker conditions. 
This approach can also be used to calculate the \emph{minimum}-entropy distribution.}. 
Hence, the asymptotic entropy per neuron of the maximum entropy model  is $\s_{isi}=\eta_2\(r_{1}\)$. The entropy-rate of the DG for $\mu=0.1$ and $\rho=0.05$ is $0.35$, and the rate of $\Delta_{hoc}=0.016$, and increases by a factor of $1.75$ if correlations increase to $0.1$. For large populations, $I_2$ of the DG can be much lower, e.g. it is  $0.57$ for $\mu=\rho=0.1$. Fig. \ref{fig:entropycomp} e also shows that the close similarity (as measured by $I_2$) between the MaxEnt-model and the DG conjectured by \cite{Macke_Berens_09} asymptotically holds for firing probabilities near $0.5$, but not necessarily otherwise.  Our results  readily  generalize to populations consisting of a finite number of homogeneous pools.   In this case, the asymptotic scaling of entropy is dominated by the within-pool correlations. Furthermore, our  results could be used to derive lower bounds on the entropy of general MaxEnt models.

\begin{figure}[t]
\includegraphics[width=\myscalerFig\linewidth]{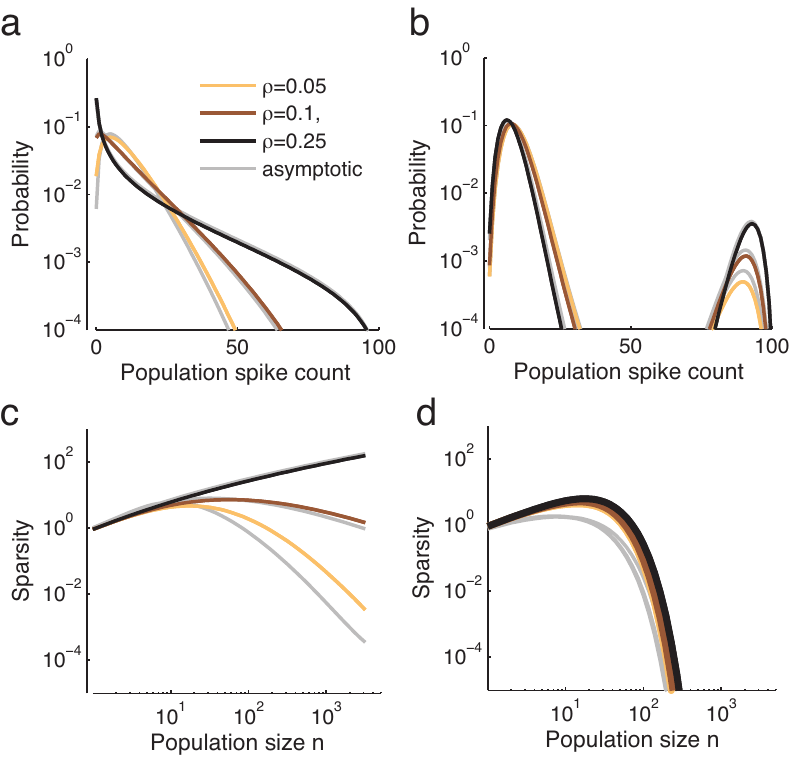}

\caption{Population spike count distributions and sparsity: {\bf a,b)} The spike count distributions for the DG (a) and Ising model (b)  for population size $n=100$ and $\mu=0.1$ (rescaled by population size $n$) are substantially different (large-n approximations in gray, background). Note that $\rho=0.25$ is the critical correlation, and that the Ising model is bimodal.
{\bf c,d)} For large $n$, the DG-population (c) is much sparser than the MaxEnt model (d), (same parameters as above). \label{fig:sparsity}}
\end{figure}

\paragraph{The hocs of the DG increase sparsity.}
In addition to the entropy, hocs also affect other population statistics. In particular, we are interested in their effect on the \emph{sparsity} of the population, which is considered to be an important feature of population coding. We quantify sparsity as the probability of the population being quiet \cite{Ohiorhenuan_Mechler_10}, i.e. P(K=0). 
It has been shown \cite{Ohiorhenuan_Mechler_10} that hocs in cortical networks lead to an increase in sparsity, and this is also consistent with the observation that MaxEnt models in the retina under-estimate the probability of quiescence \cite{Schneidman_Berry_06,Tkacik_Schneidman_09}.
We have already derived the count distribution \cite{Roudi_Aurell_09} of the DG.  From equation \eqref{eq:DGasympdistrib}, we can see that the mode of $f(r)$ is at 0, i.e.  quiescence is the most likely population state whenever the input correlation $\lambda$ exceeds the value $\lambda=0.5$ (Fig. \ref{fig:sparsity} a), which is a critical point for $f_{DG}(r)$. Interestingly, this is independent of the parameter $\gamma$ controlling the mean firing rate (as long as $\gamma<0$). For small spike probabilities $\mu$, even small correlations $\rho$ correspond to a super-critical $\lambda$   (Fig. \ref{fig:entropycomp} a).

For the corresponding Max-Ent distribution, the binary infinite range Ising model with $P(K=k)=Z^{-1}{n \choose k}\exp\(h_nk+J_nk^2\)$, we need to identify the scaling of the parameters $h_n$ and $J_n$ yielding the desired means and correlations. It should be noted that this limit is subtly, but critically different from the usual thermodynamic limit \cite{Parisi_98, Amari_Nakahara_03, Tkacik_Schneidman_09}: Scaling $J_n =J/n$ and $h_n=h$ yields a large-n distribution of $f(r) \sim \exp\(n\(\eta_e(r)+hr+Jr^2\)\)/Z$, which collapses to a single delta-peak. Thus, this approach leads to vanishing second-order correlations \cite{Amari_Nakahara_03} which violate the moment constraints. We need to ensure $(h+J)=\alpha/n$ with 
 $\alpha=(\log p_{2}-\log p_1)/(r_2-r_1)$
 to achieve correlations of order one, and this yields a large-n distribution of
\begin{align}
\eqsize
f_{isi}(r)&= Z_{isi}^{-1} \exp\(\alpha r+n \(\eta_e(r)+J(r^2-r) \) \) \label{eq:MaxEntasympdistrib}
\end{align}
with $J=(\log(r_{2})-\log(r_{1}))/(r_{2}-r_{1})$.

Figure \ref{fig:sparsity} shows a comparison of the spike count distributions of the two models for $n=100$, and the scalings of the sparsities with population size  \footnote{We assume $\rho>0$, and $r>0$, for very weak correlations, other expansions might be more accurate \cite{Roudi_Aurell_09}.}. We can see that the DG has increasing sparsity for super-critical correlation $\rho=0.25$. The count distribution of the MaxEnt model is bimodal {(corresponding to a ferromagnetic phase)}, behaves very much like a mixture of two independent distributions, and has vanishing sparsity. 
In fact, any model with interactions of finite order $q$ will asymptotically behave like a mixture of at most $q$ independent distributions \cite{Amari_Nakahara_03}, and exhibit similar sparsity scaling. Thus, correlations of all orders are necessary for achieving a continuous asymptotic spike count distribution, and the same sparsity scaling as the DG. These results were derived assuming that all neurons have identical firing rates and correlations. If the population is heterogeneous, there could be additional sparsity arising, e.g., from neurons with low firing rates. However, we conjecture that sparsity in larger populations is still strongly affected by hocs. 

\paragraph{Hocs increase heat capacity.}
Finally, we investigate the impact of hocs on the \emph{heat capacity} of the population.  As the heat capacity is proportional to the variance of log-probabilities of population states, examining it can give insights into coding properties of the population \cite{Tkacik_Schneidman_09}.  Furthermore, a sharply peaked and diverging
 specific heat (i.e. heat capacity normalized by population size) is evidence for a physical system being at a critical point \cite{Parisi_98,Stephens_Mora_08}. The distribution of a model $P(x)$ at temperature $T=1/\beta$ is given by $P_\beta(x)=P(x)^\beta/Z_\beta$, and the specific heat by $c=\mbox{Var} \log_2 P_\beta(x)/n$.  For large $n$, the spike count distribution is $P_\beta(K)=\exp(n(1-\beta)\eta_e(k/n)) P(K)^{\beta}/Z$, and asymptotically this yields

$c_{\beta}= n\int  f_\beta(r) \( \eta_2(r)^2 -\s_{\beta}^2\) dr,$
where $f_\beta$ is the limiting distribution of $P_\beta(K)$.

\begin{figure}[t]
\includegraphics[width=\myscalerFig\linewidth]{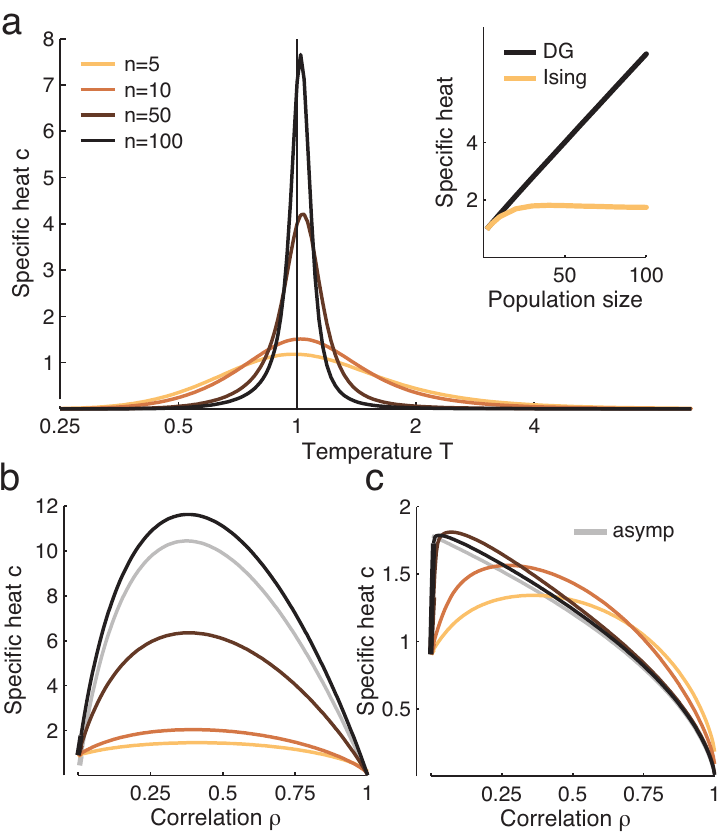}
\caption{Scaling of specific heat: {\bf a)} Specific heat of the DG (for mean $\mu=0.1$ and $\rho=0.1$) diverges at $T=1$. Inset: Specific heat of the DG at $T=1$ grows linearly with population size. {\bf b,c)} Specific heats for $\mu=0.1$ and $T=1$ vary with correlation $\rho$ for DG (b) and Ising model (c). (gray: asymptotic heat, rescaled by $100$ for DG). For large $n$, the Ising model attains it maximum at values close to $0$. 
 \label{fig:heat}} \end{figure}

Therefore, $c_\beta$ diverges linearly whenever  this integral is non-zero, which is the case for the $DG$ and many other models at $\beta=1$.  For $\beta\neq 1$, however, $f_\beta(r)$ is dominated by the exponential, collapses to  a delta-peak, and has finite specific heat. Thus, the DG has a critical point at $T=1$ (Fig. \ref{fig:heat} a). This behaviour is independent of the originally observed moments, and therefore true for almost \emph{any} such system. The second-order MaxEnt model is a notable exception, in that its $f_\beta$ consists of two symmetric delta-peaks even at $T=1$, and that its specific heat is, in general, finite for each temperature (Fig. \ref{fig:heat} a inset). Further simulations with heterogeneous all-to-all correlations  suggest that the specific heat of the $DG$ (but, in general, not of the Ising model) grows linearly in $n$ at unit temperature. 

It is therefore informative to calculate the specific heat at unit temperature as a function of the moments $\mu$ and $\rho$. In this case, the specific heat of the Ising model is
\begin{align}
\eqsize
c_{isi}&=\frac{r_1r_2J^2(\sigma+\mu^2-\mu+1/4)}{4\(1-2Jr_1r_2\)}\log_2^2(e).
\end{align}
Asymptotically, the heat capacity of the MaxEnt model is maximized for vanishing correlation, whereas the DG attains its maximum at strong correlations, e.g. $\rho=0.37$ for $\mu=0.1$   (Fig. \ref{fig:heat} b,c).
We conclude that hocs can have a substantial impact on the specific heat:  They lead to a qualitatively different scaling behaviour, and strongly influence the moments which maximize it.

\paragraph{Conclusions}
We showed that a simple binary model with common inputs could qualitatively account for a variety of empirical observations, including hocs which depend on second-order correlations, a negative strain, increased sparsity and a divergent specific heat.
It is worth remarking that all of our formulations can readily be generalized to more general input distributions or spike generation mechanisms. Further investigations will have to show whether our results would also \emph{quantitatively} account for these observations, and how they can be rigorously extended to heterogeneous and temporal correlations \cite{Burak_Lewallen_09}.

MB was supported by the Bernstein Prize (BMBF; FKZ: 01GQ0601), and JHM by a Marie Curie Fellowship.  We thank S. Gerwinn, E. Mukamel and P. Latham 
 for discussions.

\end{document}